\documentclass[useAMS,usenatbib]{mn2e}
\usepackage[latin2]{inputenc}
\usepackage[english]{babel}

\usepackage{enumerate}
\usepackage{graphicx}
\usepackage{color}
\definecolor{sz}{gray}{0.9}
\usepackage{epsfig,natbib}
\usepackage{amsmath}

\title[The stability of the terrestrial planets with a more massive "Earth"]{The stability of the terrestrial
planets with a more massive "Earth"}
\author[Áron Süli, Rudolf Dvorak and Florian Freistetter]{Áron Süli$^{1}$\thanks{E-mail:
a.suli@astro.elte.hu (ÁS)}, Rudolf Dvorak$^{2}$ and Florian Freistetter$^{3}$\\
$^{1}$Eötvös Loránd University, Department of Astronomy, Pázmány Péter sétány 1/A, Budapest, H-1518, Hungary\\
$^{2}$University of Vienna, Institute of Astronomy, Türkenschanzstrasse 17, A-1180, Vienna, Austria \\
$^{3}$Astrophysikalisches Institut und Universitäts-Sternwarte, Schillergäßchen 2-3, D-07745 Jena, Deutschland}
\begin{document}

\date{}

\pagerange{\pageref{firstpage}--\pageref{lastpage}} \pubyear{2005}

\maketitle

\label{firstpage}

\begin{abstract}
Although the long-term numerical integrations of planetary orbits indicate that our planetary
system is dynamically stable at least $\pm$4 Gyr, the dynamics of our Solar System includes
both chaotic and stable motions: the large planets exhibit remarkable stability
on gigayear timescales, while the subsystem of the terrestrial planets is weekly chaotic 
with a maximum Lyapunov exponent reaching the value of 1/5 Myr$^{-1}$. In this paper
the dynamics of the Sun--Venus--Earth--Mars-Jupiter--Saturn model is studied, where
the mass of Earth was magnified via a mass factor $\kappa_E$. The resulting systems
dominated by a massive Earth may serve also as models for exoplanetary systems that 
are similar to our one. This work is a continuation of our previous study, where the 
same model was used and the masses of the inner planets were uniformly magnified. That 
model was found to be substantially stable against the mass growth. Our simulations were 
undertaken for more then 100 different values of $\kappa_E$ for a time of 20,
in some cases for 100 Myrs. A major result was the appearance of an instability window 
at $\kappa_E \approx 5$, where Mars escaped. This new result has important implications 
for the theories of the planetary system formation process and mechanism. It is shown that with 
increasing $\kappa_E$ the system splits into two, well separated subsystems: one consists of 
the inner, the other one consists of the outer planets. According to the results the model 
became more stable as $\kappa_E$ increases and only when $\kappa_E \ge$ 540 Mars escaped, on a
Myr timescale. We found an interesting protection mechanism for Venus. These results give 
insights also to the stability of the habitable zone of exoplanetary systems, which harbour 
planets with relatively small eccentricities and inclinations.
\end{abstract}


\section{Introduction}

The determination of the stability of our Solar System is one
of the oldest problems in astronomy. The question has been debated
over more than 300 years, and has attracted the attention of many
famous mathematicians over the course of history. The problem
played a central role in the development of non--linear dynamics
and chaos theory. Despite the considerable efforts, we do not possess
a definite answer to the question of whether our Solar System is stable or
not. This is partly a result of the fact that the definition of the term
stability is not unambiguous when it is used in relation to the problem
of planetary motion. In addition to the vagueness of the concept of stability, the
planets in our planetary system show a character typical of dynamical chaos.
The physical basis of this chaotic behaviour is now partly understood as a consequence of
resonance overlapping and three body resonances \citep{Lecar2001,Murray99,Nesvorny99}
which can manifest themselves in dramatic and relatively sudden changes in an orbit. 
In the last two decades several numerical stability studies were performed in order to 
throw light on the question. At present the longest numerical integrations published are 
those of \cite{Ito2002}, where six long-term numerical integrations of all nine planets, 
covering a time-span of several $10^9$ and $10^{11}$ years are discussed. Their fundamental 
conclusion is that the Solar System seems to be stable in terms of the Hill-criteria at least 
over a time-span of $\pm$4 Gyr. Moreover it turned out that during the integration period all 
the planetary orbital elements have been confined in a narrow region.

On the other hand according to Laskar's semi-analytical secular perturbation theory
\citep{Laskar88}, the terrestrial planets', especially Mercury's and Mars' eccentricities and
inclinations show large and irregular variations on a time-scale of several $10^9$ year 
\citep{Laskar96}.

\begin{figure}
\centerline{
\includegraphics[width=1.0\linewidth,angle=0]{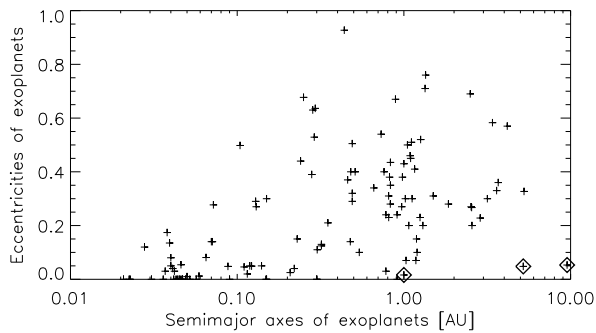}}
\caption[]{Eccentricities vs. the semimajor axes of the observed extrasolar planets. The
$x$-axis is logarithmic. The positions of the Earth, Jupiter and Saturn
are also indicated as diamonds with a plus sign in the middle.}
\label{fig1}
\end{figure}

Nowadays to study the stability of the Solar System, or its variants, as a representative 
of the different planetary systems has become part of the frontline research. Over the 
past few years the detection of planets outside the Solar System, the so called exoplanets, 
has greatly stimulated the stability studies of planetary systems. New exoplanets are being 
discovered on a regular basis; more than 150 (April, 2005) exoplanets
are now known. There are 136 systems consisting of a central star and a gaseous giant planet, 
and 14 multiple systems with two, three and four planets. The so far discovered exoplanets 
have a minimum mass range ($m\cdot \sin(i_p)$) from 0.042 $m_J$ to 17.5 $m_J$ where $m_J$ is 
Jupiter's mass and $i_p$ is the inclination of the orbital plane with respect to the plane of 
the sky. Since $i_p$ is unknown a precise mass cannot be determined, only a lower mass limit.
Because of the technical limitations only planets of Neptune mass or above can be detected and 
then only if they are less than 5 AU or so from the star 
\footnote{With OGLE, which is based on optical gravitational lensing it is possible to detect 
exoplanets with only Earth-mass.}. 
Therefore more than 90 \% of these planets are orbiting their host star well inside Jupiter's
orbit. There are major differences between the characteristics of the so far observed systems and 
those of the Solar System. Most of the planets have minimum masses substantially greater than that
of Jupiter -- up to six or even more times the mass of Jupiter. Dozens of planets are orbiting very
close to their hosting star, with semimajor axes down to 0.04 AU. Finally, planetary orbits are 
found with large eccentricities, up to approximately 0.7, plus a few greater, significantly greater 
than the highest 
eccentricities observed for planets in our Solar System. These characteristics are depicted 
in Fig. \ref{fig1}, where the planets' eccentricities are plotted against their semimajor 
axes, the locations of the Earth, Jupiter and Saturn are marked as diamonds. From 
Fig. \ref{fig1} it is apparent that our planetary system may serve as model case for those 
planetary systems which have small eccentricities. Presumably these are also the ones where we 
may expect stable terrestrial planets moving in habitable zones \citep{Asghari2004}.

In previous papers \citep{Dvorak2002, Dvorak2005} the dynamical evolution of a simplified Solar 
system was studied. The model consisted of the Sun, the three most massive terrestrial planets
(Venus, Earth, Mars), Jupiter and Saturn. The masses of the inner planets were uniformly 
magnified by a mass factor $\kappa$. It turned out that the different systems remained stable 
up to 10 Myr for $\kappa \le 220$. Stable islands were found for $\kappa$ = 245 and 250, which 
is a well-known property in such regions which are close to the last stable orbit in the chaotic 
domain. We have shown that the dynamical coupling of Venus and Earth and that of Jupiter and 
Saturn remained unbroken for all studied $\kappa$. On the other hand the motion of Mars was not 
coupled to any other planet, what may be a reason for the fact that it was always Mars which caused
the decay of the system after close approaches with Earth. However the remarkable stability of 
these model planetary systems suggests that exoplanetary systems with configuration like our 
Solar System may harbour moderately or even very massive terrestrial-like planets.

In the present work our aim was to study and analyze the dynamical evolution and the stability 
of the system with respect to the masses involved. Contrary to our previous study, in the present 
work only the mass of the Earth was increased via a mass factor $\kappa_E$. Furthermore, the 
examined systems may be considered as models for individual exoplanetary system, and the results 
can be applied to them. Section 2 explains our dynamical model, the applied methods and section 3 
is devoted to a detailed description of the results in the $\kappa_E \in [1,600]$ region. Finally, 
we discuss the results and the implication for exoplanetary systems.

\section{Description of the model and methods}

\begin{figure*}
\begin{minipage}{120mm}
\centerline{
\includegraphics[width=0.9\linewidth,angle=0]{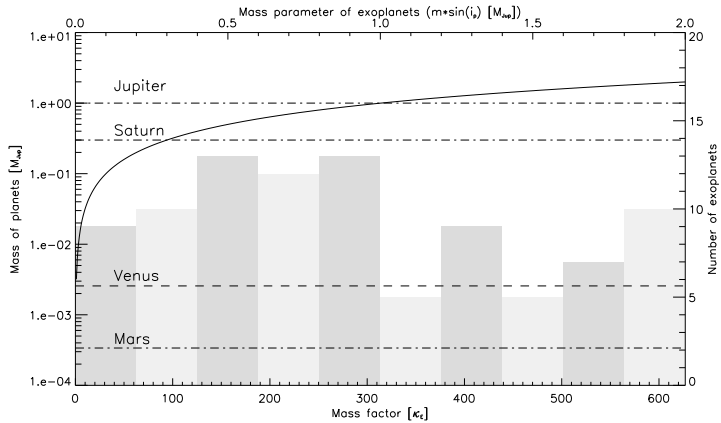}}
\caption[]{The mass distribution of the known exoplanetary systems are plotted together with the 
mass of the Earth as a function of the mass factor ($\kappa_E$) (solid line). The masses of the 
other planets were left unchanged, and they are represented by straight, dashed lines. The lower 
$x$-axis is the mass factor, the upper $x$-axis is the mass parameter of the exoplanetary system.
The left $y$-axis is the mass of the planets in a logarithmic scale, the right $y$-axis is the 
number of planets. The width of the bin is 0.2 (this refers to the top axis)}
\label{fig2}
\end{minipage}
\end{figure*}

The applied dynamical model consisted of the Sun, Venus, Earth, Mars, Jupiter and Saturn.
We have chosen this model for two reasons. To speed up the numerical integrations, we have
omitted Mercury, Uranus and Neptune. Because of the small mass of Mercury, it only slightly
perturbs the motion of the terrestrial planets and it has a short orbital period which
would require a reduction in the integration step-size resulting in increased CPU time.
Even though Uranus and Neptune are massive
planets, they are evolving around the Sun more than twice as far as Saturn, and so they do 
not influence the motion of the inner planets significantly. These simplifications are also 
justified by the fact, that the dynamics of the inner planets are dominated by Jupiter and 
Saturn. Furthermore the so far observed exosystems harbour at most three of four planets. 
The above modification of the Solar System gives grounds for the adjective "simplified". 
Hereafter we will refer to our simplified Solar System as $S^3$. The other reason was to 
make the model more like an exoplanetary system. The mass of the Earth was magnified via a 
mass factor $\kappa_E$ (the masses of the other planets were unaltered), which resulted in 
such hypothetical planetary systems, whose characteristics agree very well with several 
exosystems (for example if we multiply the semimajor axes of the planets of 47 Ursae Majoris 
by 2.5, the resulting configuration is something like the Sun, Jupiter and Saturn system, if
$\kappa_E$=1). According 
to these modifications, our models are parallel with those exosystems, which harbour planets 
with small eccentricities and inclinations. In Fig. \ref{fig2} the masses of the planets
are plotted as a function of $\kappa_E$. At $\kappa_E \approx 90$ Earth is as massive as 
Saturn and when $\kappa_E \approx 300$ Earth is as massive as Jupiter. The mass of the other
planets are plotted by horizontal dashed lines. The mass distribution of the so far discovered 
exoplanetary systems is also depicted in Fig. \ref{fig2} up to mass parameter = 2. The 
integrations were done in the $\kappa_E \in [1,600]$ region, which approximately corresponds
to 0 $<$ mass parameter $\approx$ 2 interval, and therefore the mass distribution in Fig.
\ref{fig2} is displayed only in this interval.

We have considered the $S^3$ as a non-linear Hamiltonian system, governed only by classical 
Newtonian gravitational forces between the objects of the model. The planets have been taken 
as point masses, and the Earth's Moon was not included in the models. The initial planetary 
orbital elements and the actual masses are listed in Table \ref{tab:2.1}.

To check whether the different $S^3$ setups belonging to different $\kappa_E$s could be stable over a 
long time interval we utilized the very precise numerical integration scheme, the 
Lie-integrator method. This method is based on the integration of differential equations with 
Lie-series and uses the property of recurrence formulae for the Lie-terms. The details of the 
method are described in \cite{Hanslmeier1984} and \cite{Lichtenegger84}. The scheme is particularly 
effective in the case of highly eccentric orbits. The accuracy of this integration technique 
is based on an automatic stepsize control, and it has been checked in several comparative test 
computations with other integrators. Although symplectic integrators are very effective when 
eccentricities remain small, but the Lie-integrator is a better choice in studies, where very large 
eccentricity orbits may occur.

The length of the integrations was fixed at 20 Myrs, in some cases at 100 Myrs, which was a trade-off 
between too long CPU time and the quality of the results. Since our interest focuses primarily on 
the inner three planets' motions, for which the orbital time-scales are much shorter than those of 
the outer two giant planets, the 20 Myrs seems a justifiable choice, although it is known from earlier 
work (see for e.g. \cite{Jones2001,Jones2002}) that exosystems can stay stable for hundreds of 
millions of years and then fall apart. In the case of weak chaos therefore the 100 and especially the 
20 Myr timespan might be short and in some of the integrations where the terrestrial planets survived 
for 20 or 100 Myrs, they might not survive significantly longer. For the sake of a comparison 
all results were derived on the same computer. In some cases we have performed comparative integrations
on different platforms (to obtain more information of the particular system).

The conservative definition of the point at which systems become unstable is when close encounter between 
two planets happens: two bodies approach one another within an area of the larger Hill radius. The consequence 
of such an event is the dramatic changes in the orbital elements of the two planets, and usually the escape
of the planet with smaller mass. In this paper we define a system unstable, when an orbit crossing or a close
encounter happens. This definition is somewhat more general and the instability can be directly connected to 
the eccentricity via the perihelia and aphelia distances. Henceforward we state that our model is dynamically 
unstable if orbit crossing or close 
encounters happen in the course of the integration. Using both criteria is clearly an extension of the 
conservative definition of instability. It is justifiable to incorporate the orbital crossing criteria in the 
definition since we know from experience that an orbital crossing in general leads to a close encounter 
in very short time. The main difference between the two definitions is the time-scale of instability. We note 
that orbital crossing does not lead to close encounter in all cases when certain resonances have adjusted the 
planetary motions in such a way that the planets avoid each other. For example this is the case in the 
Neptune-Pluto pair. In our models no such protection resonances are present, henceforward we use the above 
definition to distinguish between stable and unstable $S^3$ setups. The integrations were not stopped after
one of the above criterion had been met, but were continued until the integration timespan was reached.

\begin{table*}
\centering
\begin{minipage}{140mm}
\caption[]{Planetary orbital elements (JD 2449200.5) with respect to the mean
ecliptic and equinox J2000. The quantities $a$, $e$, $i$, $\omega$, $\Omega$ and $M$ denote the
semimajor axis, eccentricity, inclination, argument of perihelion, longitude of ascending node and
mean anomaly. In the last row the masses of the planets in Solar mass units are listed ($M_{Sun}=1$).}
\center
\begin{tabular}{lrrrrr}
\hline
  & Venus & Earth & Mars & Jupiter & Saturn\\
\hline
$a$ &  0.723328  &  0.999999 &  1.523614  &  5.202627 &  9.545509\\
$e$ &  0.006747  &  0.016716 &  0.093443  &  0.048370 &  0.052420\\
$i$     &  3.394820  &  0.000545 &  1.850191  &  1.304638 &  2.485620\\
$\omega$ & 54.847892  &113.611521 &286.492727  &275.222227 &338.025839\\
$\Omega$ & 76.691772  &349.288391 & 49.573832  &100.470086 &113.651098\\
$M$      &135.521541  & 78.172620 &185.208769  &233.733076 &259.852365\\
$1/m_i $ &  408 523.71 &  332 946.047  &  3 098 708.0  &  1047.348 &  3497.898\\
\hline
\end{tabular}
\label{tab:2.1}
\end{minipage}
\end{table*}

For an indication of stability we used a straightforward check based on the eccentricity. This 
osculating orbital element shows the probability of orbital crossing and close encounter of two 
planets, and therefore 
its value provides information on the stability of the orbit. We examined the behavior of the 
eccentricities of the planets along the integration, and used the largest value as a stability 
indicator; in the following we call it the maximum eccentricity method (hereafter MEM). This is 
a reliable indicator of chaos, because the overlap of two or more resonances induce chaos and 
large excursions in the eccentricity. We know from experience, that instability comes from a chaotic 
growth of the eccentricity. This simple check has already been used in other stability studies, 
and was found to be quite a powerful indicator of the stability character of an orbit 
\citep{Dvorak2003,Asghari2004}.

\subsection{The Laplace-Lagrange secular theory}

In order to find a theoretical explanation for the decay of the system we have applied the 
Laplace-Lagrange first order secular theory. This linear theory yields accurate results under the following 
assumptions:
\begin{enumerate}
\item no mean motion commensurabilities,
\item no orbit-crossing, and
\item the eccentricities and inclinations are small enough.
\end{enumerate}
Our models meet these criteria. However a complication is that the orbits of Jupiter and Saturn
are close to a 5:2 commensurability. Since the appearance of the Laplace-Lagrange theory, as a 
first approximation it has been extensively used in the studies of motion of the planets and of
other Solar System bodies. In several researches \citep{Knezevic86,Laskar88} the results of the 
first and higher order secular theories were compared. According to these studies the secular 
frequencies calculated from the Laplace-Lagrange theory are sufficiently accurate for our present 
research goal. The main discrepancies are in the case of Jupiter and Saturn. As we are interested 
only in the motions of the inner planets, therefore we will apply the Laplace-Lagrange theory to 
our models. From the above assumptions it is apparent that the precision of the theory does not 
depend on the planetary masses, accordingly there is no theoretical limitation on the mass factor's 
magnitude.

Since the eccentricities and inclinations may vanish at remote epochs, it is better to use
the Lagrange orbital elements:
\begin{eqnarray}
\left( \begin{array}{c}
h \\
k \\
\end{array} \right) =
e \cdot \begin{array}{c}
\sin \varpi \\
\cos \varpi \\
\end{array}
,\qquad
\left( \begin{array}{c}
p \\
q \\
\end{array} \right) =
i\cdot \begin{array}{c}
\sin \Omega \\
\cos \Omega \\
\end{array}
. \label{eq:2.1}
\end{eqnarray}
Eq. (\ref{eq:2.1}) associates the $h$, $k$, $p$ and $q$ Lagrangian-elements to the $e$, $i$, $\varpi$, 
and $\Omega$ Keplerian-elements, where $e$ denote the eccentricity, $i$ the inclination, $\varpi$
the longitude of perihelion and $\Omega$ the longitude of ascending node. Using these variables the 
general solution of the differential equations for the planets takes the following form:
\begin{eqnarray}
\left( \begin{array}{c}
h_s \\
k_s \\
\end{array} \right) &=&
\sum_{j=1}^n M_s^{(j)} \begin{array}{c}
\sin \\
\cos \\
\end{array}
\left( g_jt+\beta_j\right), \label{eq:2.2} \\
\left( \begin{array}{c}
p_s \\
q_s \\
\end{array} \right) &=&
\sum_{j=1}^n L_s^{(j)} \begin{array}{c}
\sin \\
\cos \\
\end{array}
\left( f_jt+\gamma_j\right), \label{eq:2.3}
\end{eqnarray}
where the $s$ index denotes the planet, the $j$ index denotes the mode, $N$ is the number of 
planets, $M_s^{(j)}$ and $L_s^{(j)}$ are the amplitudes, $g_j$ and $f_j$ denote the secular 
frequencies and $\beta_j$ and $\gamma_j$ are the angular phases.

The planet's orbital elements are described by Eq. (\ref{eq:2.2}) and Eq. (\ref{eq:2.3}), which 
are the sum of harmonic oscillations. Using these formulae it can be calculated that the planet's 
eccentricities and inclinations are varying between given limits with quasiperiodic oscillations. 
Due to the positive $g_j$ secular angular velocities the apsidal lines of the planets are rotating 
directly, whereas the nodes accordingly to the negative $f_j$ secular angular velocities are rotating 
indirectly. Upon these mean rotations quasiperiodic variations are superimposed. Both of the apsidal and 
nodal motions can be approximated by average angular velocities, which are to a first approximation
equal with the frequencies of those harmonious terms which are multiplied by the largest amplitudes:
\begin{eqnarray}
e_s \cdot \begin{array}{c}
\sin \varpi_s\\
\cos \varpi_s\\
\end{array}
\approx 
M_s^{(J)} \begin{array}{c}
\sin \\
\cos \\
\end{array}
\left( g_Jt+\beta_J\right) \label{eq:2.4}
\end{eqnarray}
where $M_s^{(J)} = \mathbf{max}_j |M_s^{(j)}|$ and the average angular velocity of the $s$'th planet is 
given by $g_J$.

In this manner on the basis of the amplitudes each secular frequency can be associate with each planet. 
This association is not mutually unambiguous. It may happen to associate a certain secular frequency to 
more then one planet.

The use of the linear theory seems to be in contradiction to the large eccentricities and inclinations 
that may be reached by a planet during the simulation. We emphasize that the forementioned theory
was used at the beginning of the integration, when the inclinations and the eccentricities are small,
and the above assumptions are therefore fullfilled.

The above described linear secular theory was implemented in the {\sc MAPLE} computer algebra program. 
With the aid of this application the formulae of the theory can be evaluated in a few seconds and it 
gives the complete first order solution of the problem.

\section{Results}

In this section we give a detailed description and overview of the results of our simulations. 
The model was integrated for more then 100 different values of $\kappa_E$. The interval of data 
output was 100 years, the total amount of data is approximately 6 GBs, and the total used CPU time 
is more than several thousands of days. All of the integrations were done on two Sun Fire 15000
supercomputers with 72 US-III+ 1200 MHz processor in each computer.
In Table \ref{tab:3.1} the first column lists the different 
$\kappa_E$ intervals, the second shows the stepsize ($\Delta \kappa_E$) in the mass factor. A small 
stepsize is taken for $4 \le \kappa_E \le 6$ in order to explore the interesting behaviour of the 
system leading -- very surprisingly -- to the escape of Mars. In the third column we list 
approximately the mass of Earth in Jupiter's mass unit. The last column gives the time-span 
of numerical integrations.

\begin{table}
\center
\caption[]{Summary of the integrations.}
\begin{tabular}{crrrrc}
\hline
$\kappa_E$ interval   & $\Delta \kappa_E$  & Earth mass in & Total time \\
 & & Jupiter mass & [Myr] \\
\hline
1 -- 25      &  1.0              &  [1/300 -- 1/12]  &  20 \\
4 -- 6       &  0.1              &  [1/75 -- 1/50]   & 100 \\
30 -- 200    &  5.0              &  [1/10 -- 2/3]    &  20 \\
210 -- 300   & 10.0              &  [2/3 -- 1]       &  20 \\
330 -- 600   & 30.0              &  [1 -- 2]         &  20 \\
\hline
\end{tabular}
\label{tab:3.1}
\end{table}

\subsection{The $1 \leq \kappa_E \leq 25$ region}

As we expected, the orbital motions of the planets indicate long-term stability in most of
our numerical experiments: no orbital crossings nor close encounters between any pair of
planets took place in the course of the integrations. However, a suprising result was the
discovery of the instability window for the $\kappa_E \in [4,6]$ interval: at several 
$\kappa_E$ values Mars escaped; the details can be found in the next subsection.

\begin{figure}
\centerline{
\includegraphics[width=1.0\linewidth,angle=0]{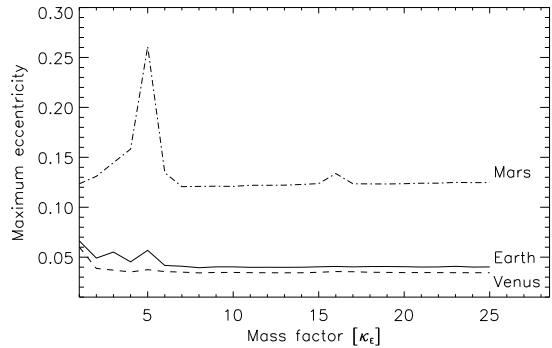}}
\caption{Results of the MEM in the $1 \le \kappa_E \le 25$ region ($\Delta \kappa_E = 1$).
The $x$-axis is the $\kappa_E$, the $y$-axis is the maximum eccentricity (ME).}
\label{fig3}
\end{figure}

Fig. \ref{fig3} depicts the results of the MEM. The maximum eccentricities (hereafter ME)
of Jupiter and Saturn are actually constant, with a value of 0.06, and 0.088, respectively 
(they are not shown in Fig. \ref{fig3}). As one can clearly see from Fig. \ref{fig3}
the MEs of the Earth and Venus are relatively small, and both curves show a similar behaviour 
as a consequence of the well-known coupling between them. We note, that the ME of Earth for 
$\kappa_E=1$ is 60\% greater than for $\kappa_E=2$. After some oscillations of Venus' and 
Earth's MEs, they stay almost constant with a value of 0.041 and 0.035, respectively. In 
turn, the ME of Mars steadily increases with $\kappa_E$, and at $\kappa_E=5$ it suddenly 
reaches a very high value, $e_{Mars}=0.26091$ (the perihelion distance of Mars is $q_{Mars}
 = 1.126$)! After this peak Mars' ME drops down to its starting value 0.1231, and begins to 
 increase slowly and gradually with $\kappa_E$.

\begin{figure*}
\centering
\includegraphics[width=0.7\linewidth,angle=0]{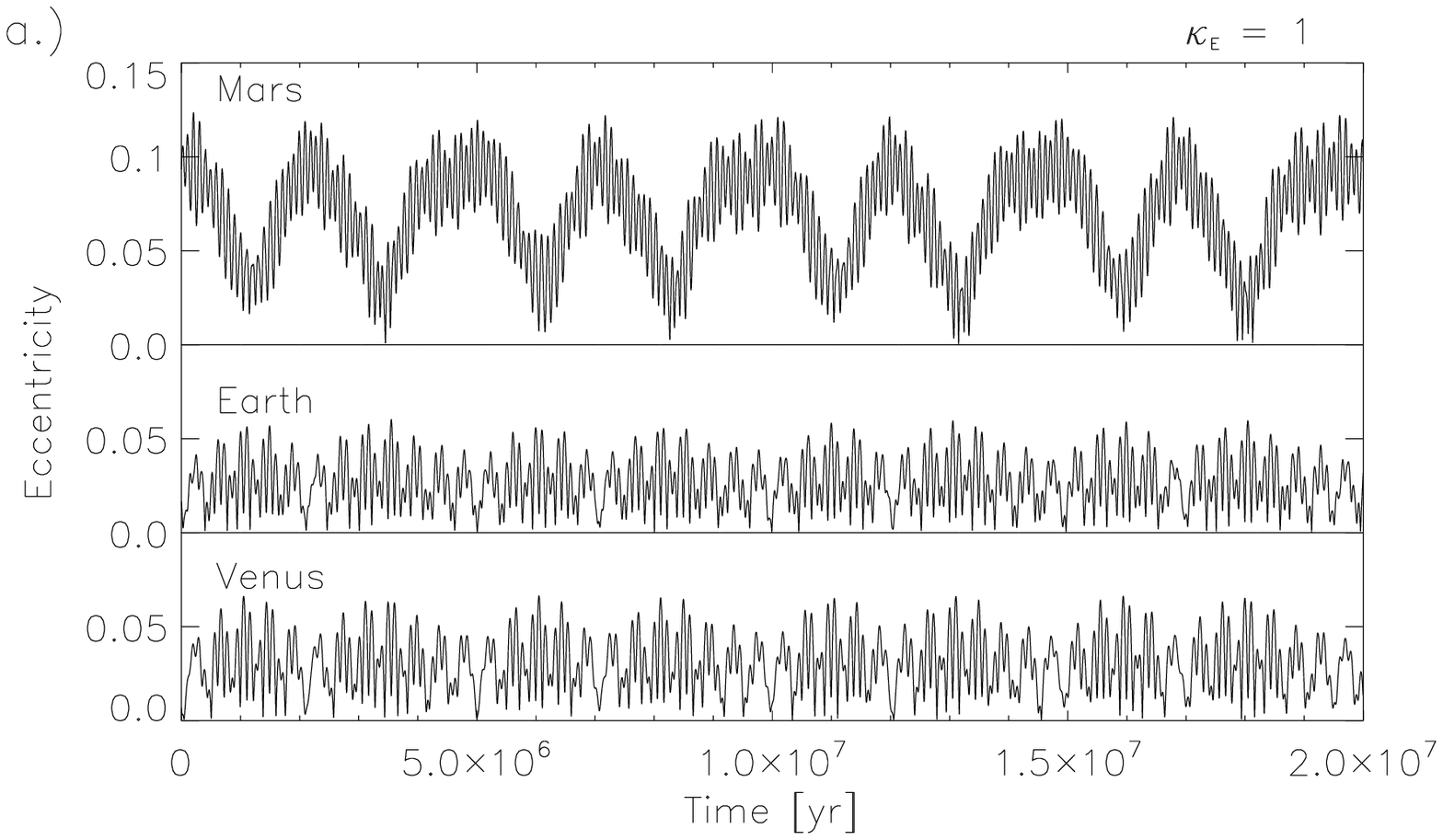}
\includegraphics[width=0.7\linewidth,angle=0]{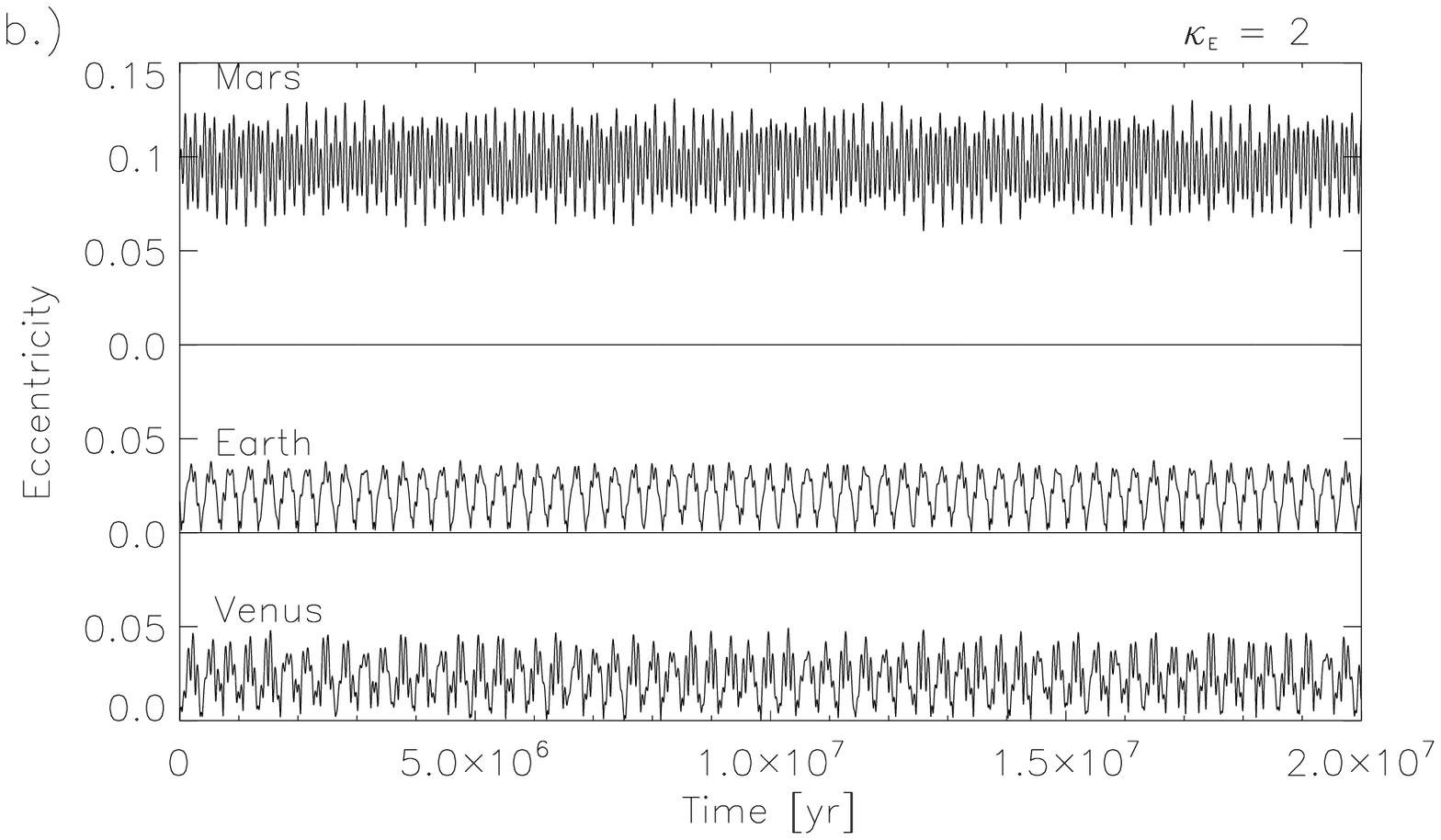}
\includegraphics[width=0.7\linewidth,angle=0]{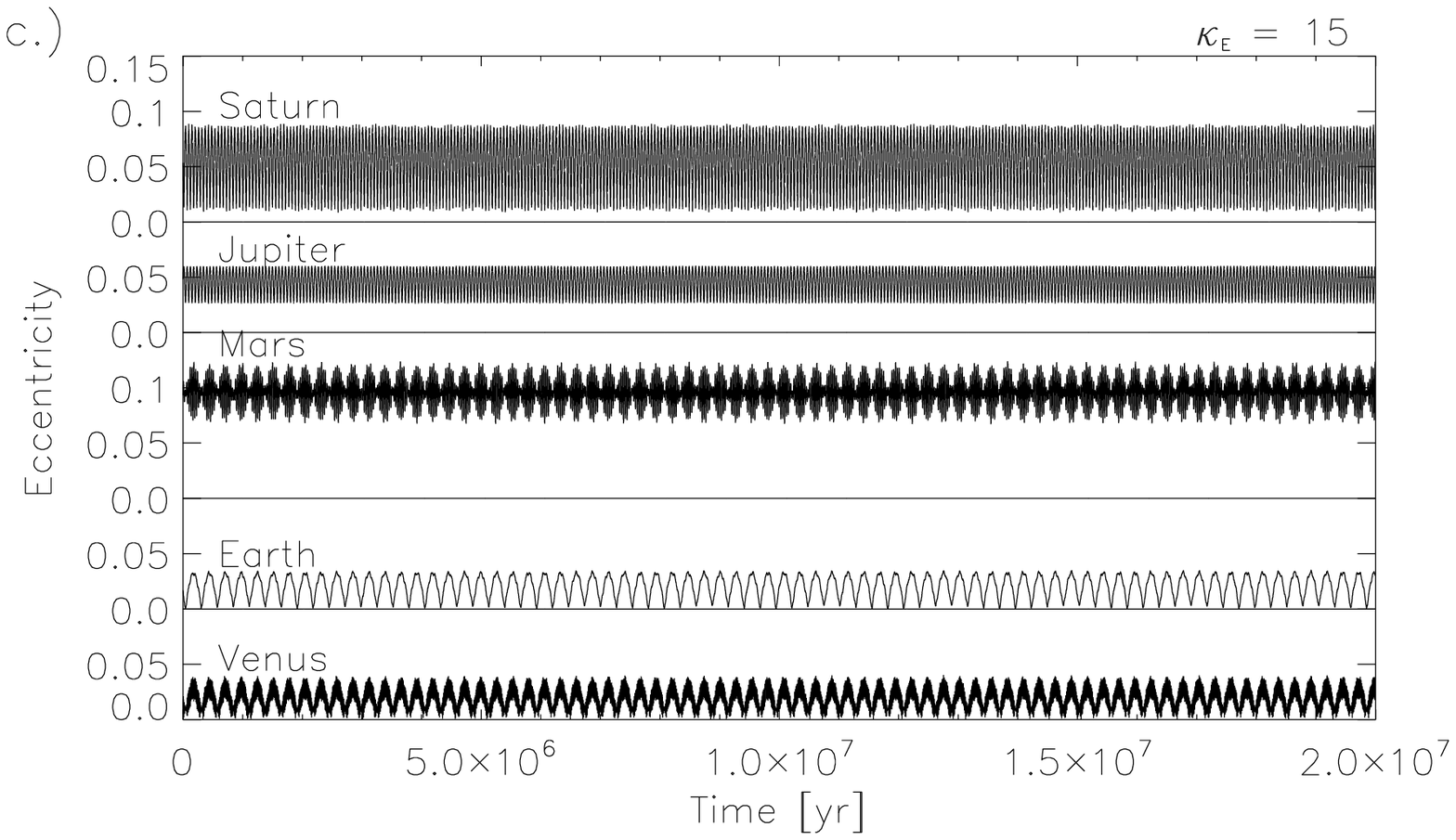}
\caption{The eccentricities of the inner planets are shown as a function of time for the 
$\kappa_E=1,\,2,\,15$ simulations. The time and mass factor evolution of the planets' eccentricities 
can be traced on the panels. On panel 4c. also Jupiter's and Saturn's eccentricities are shown: 
the existence of the two separate subsystems is well visible.}
\label{fig4}
\end{figure*}

The character of the variation of the planetary orbital elements does not change substantially 
over the course of the simulation, except for $\kappa_E$ around 5. The variations of the 
semimajor axes are in the order of $10^{-4}$--$10^{-3}$ AU for the inner planets, and for 
Jupiter and Saturn they are approximately $10^{-3}$, $10^{-2}$ AU, respectively. The variations 
in Jupiter's and Saturn's semimajor axes are one order of magnitude larger than those of the 
inner planets, which is a consequence of the 5:2 near commensurability between their motion.
For all simulations in this mass factor region the behaviour of the semimajor axes did not change,
and their excursions remained in the above range.

The eccentricities as a function of time of the inner planets for $\kappa_E$ = 1, 2, and 15
are shown in Fig. \ref{fig4} (for the sake of a comparison the eccentricities for the actual
masses are also depicted). As $\kappa_E$ goes from 1 to 2, there is a substantial change in $e$ 
for all three terrestrial planets. According to the MEM analysis (see Fig. \ref{fig3}), the 
MEs of Earth and Venus decrease, which manifests themselves in greater perihelia and smaller 
aphelia distances (see Fig. \ref{fig4}b). These changes decrease the probability of a close 
encounter between Venus and Earth, as can be inferred form Fig. \ref{fig4}b. Beyond this $\kappa_E$ 
value the behaviour of 
Earth's and Venus' eccentricities do not change and in general their orbits turns into a more 
and more regular, quasiperiodic ones as $\kappa_E$ increases. Moreover their dynamical coupling 
strengthens. These features can be traced on the panels of Fig. \ref{fig4}.

In the case of Mars the character of the variation of the eccentricity is more complex:
\begin{itemize}
\item For the actual masses Mars' eccentricity fluctuates with an amplitude of 0.1 and with a
very long period of approximately 2.5 Myrs. To this fluctuation smaller amplitudes with shorter
periods are added. It is clear from Fig. \ref{fig4}a that Mars' motion is not coupled to any 
of the other two terrestrial planets.
\item At $\kappa_E$ = 2 Mars' eccentricity shows a quite different behaviour: it oscillates with 
a very short period around a mean eccentricity of 0.1 with an amplitude of about 0.02. With 
increasing $\kappa_E$ the 
center of oscillation shifts to higher values, and reaches its maximum at 5, where the system is 
destabilized.
\item For $\kappa_E > 5$ the motion of Mars gradually turns into a regular one. Moreover, dynamical 
coupling develops between all the three inner bodies, which is very well visible from Fig. \ref{fig4}c.
\end{itemize}
The terrestrial planets seem to gradually form a subsystem as the mass factor increases. As the Earth 
becomes the dominant planet in this region and the motions of the two smaller neighboring planets are 
influenced primarily by the Earth (the major variation in the eccentricities of the Earth, Venus 
and Mars has the same period, see Fig. \ref{fig4}c). With regard to this we may consider the inner 
planets as a collection of dynamically mutually dependent planets, namely a subsystem.

In this mass factor region the giant planets move on quiet orbits for the duration of all numerical 
integrations. Jupiter and Saturn also constitute a subsystem which is practically not affected from 
the terrestrial planets. An analysis of the eccentricities of all bodies for the subsequent mass 
factors shows that the model graduates into a system which consists of two separate and loosely 
dependent subsystems.

\subsection{The instability window at $\kappa_E \approx 5$}

\begin{figure}
\centering
\includegraphics[width=1.0\linewidth,angle=0]{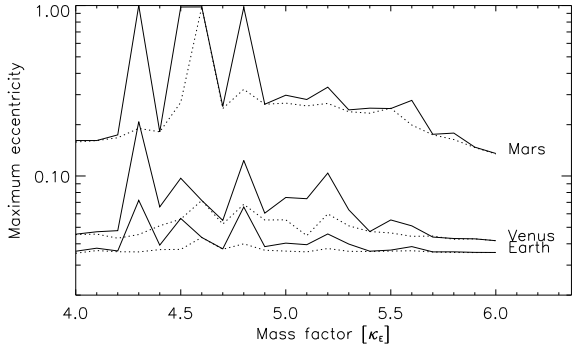}
\caption{Results of the MEM in the $4 \le \kappa_E \le 6$ region. The dotted lines show the ME of 
the 20 Myr, whereas the solid lines show the ME of the 100 Myr integrations. The MEs of Jupiter and 
Saturn are not plotted. The instability window occupies the (4.2, 4.9) region. We note that the 
ordinate is logarithmic.}
\label{fig5}
\end{figure}

According to Fig. \ref{fig3} the ME of Mars reached its maximum value at $\kappa_E$ = 5. To explore 
the dynamical evolution of the inner planets in more detail simulations were performed in the $\kappa_E$ 
= [4,6] region, with a stepsize of 0.1 for 100 Myrs. The results of the MEM are shown in Fig. \ref{fig5},
where several peaks ($e_M = 1.0$) in the curve of Mars' ME are visible, which correspond to escape 
orbits. We note that there are two minima: at 4.4 and at 4.7. This feature is typical of chaotic 
systems, where small differences in initial conditions, round off errors or the applied computing 
architectures may result in different outcomes. To verify this, the two systems with $\kappa_E$ = 4.4 
and 4.7 were integrated on different computers and both of the systems decayed. After $\kappa_E = 5.2$ 
the ME of Mars gradually drops down and at $\kappa_E = 6$ it is already less than 0.15.

\begin{figure}
\centering
\includegraphics[width=1.0\linewidth,angle=0]{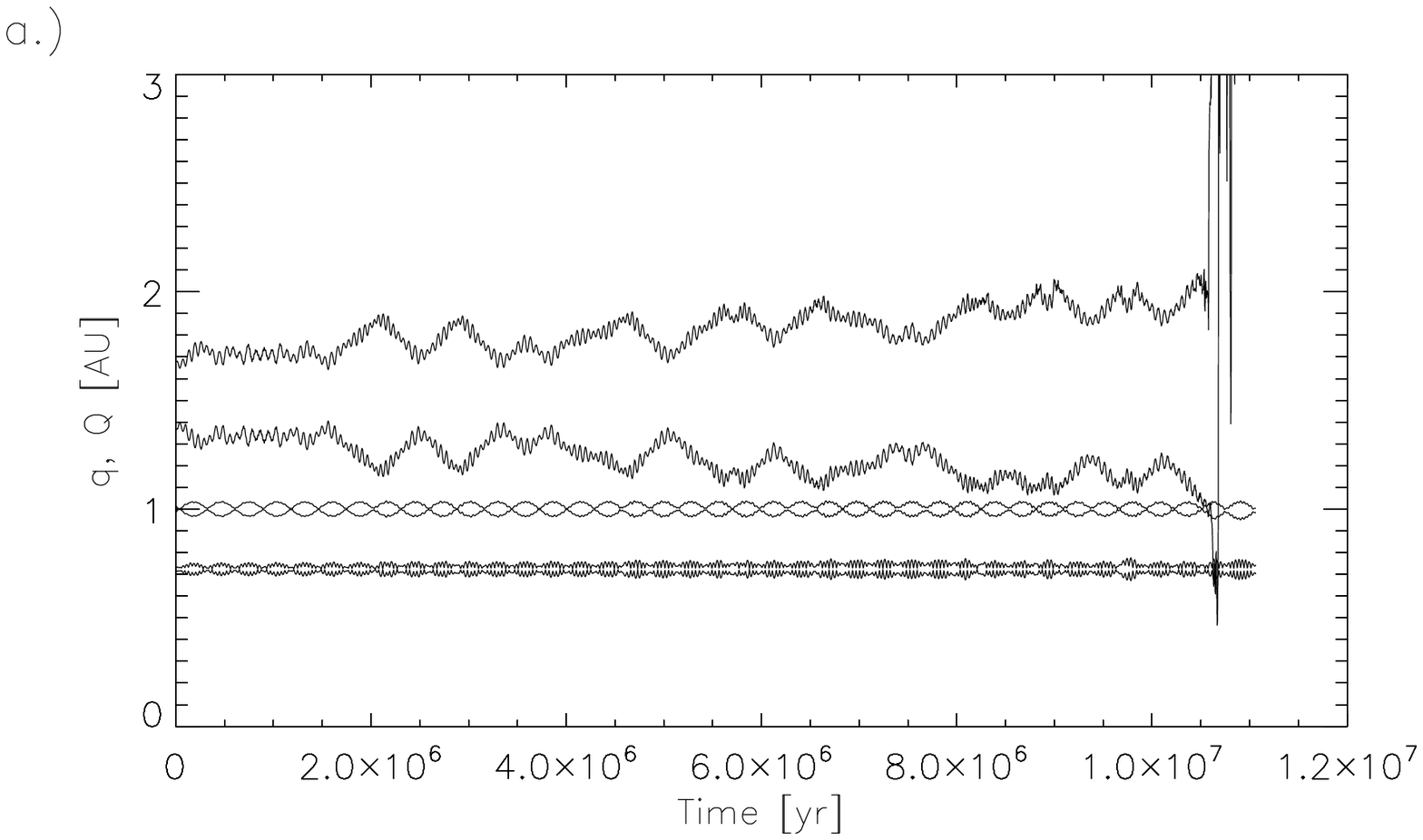}
\includegraphics[width=1.0\linewidth,angle=0]{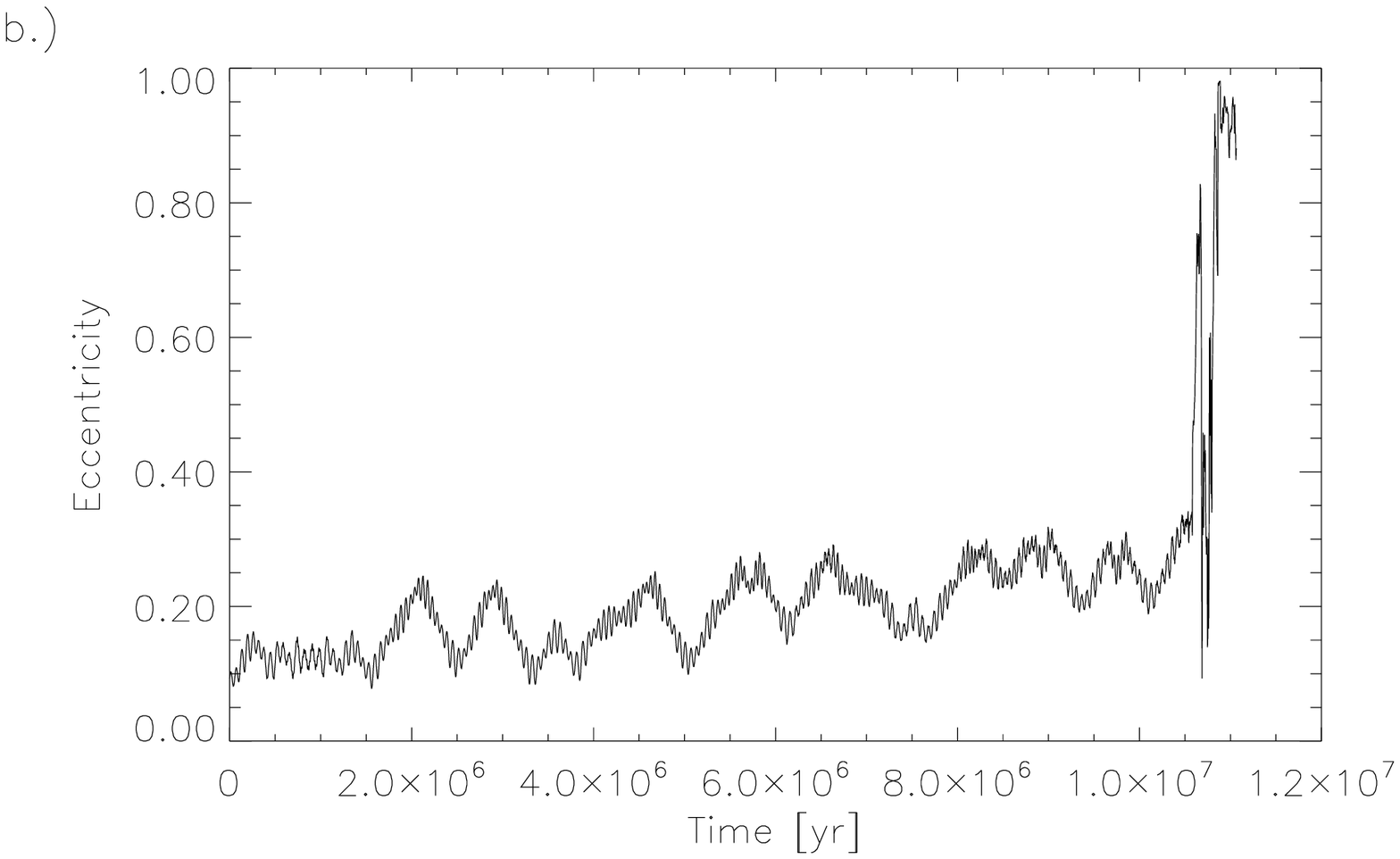}
\includegraphics[width=1.0\linewidth,angle=0]{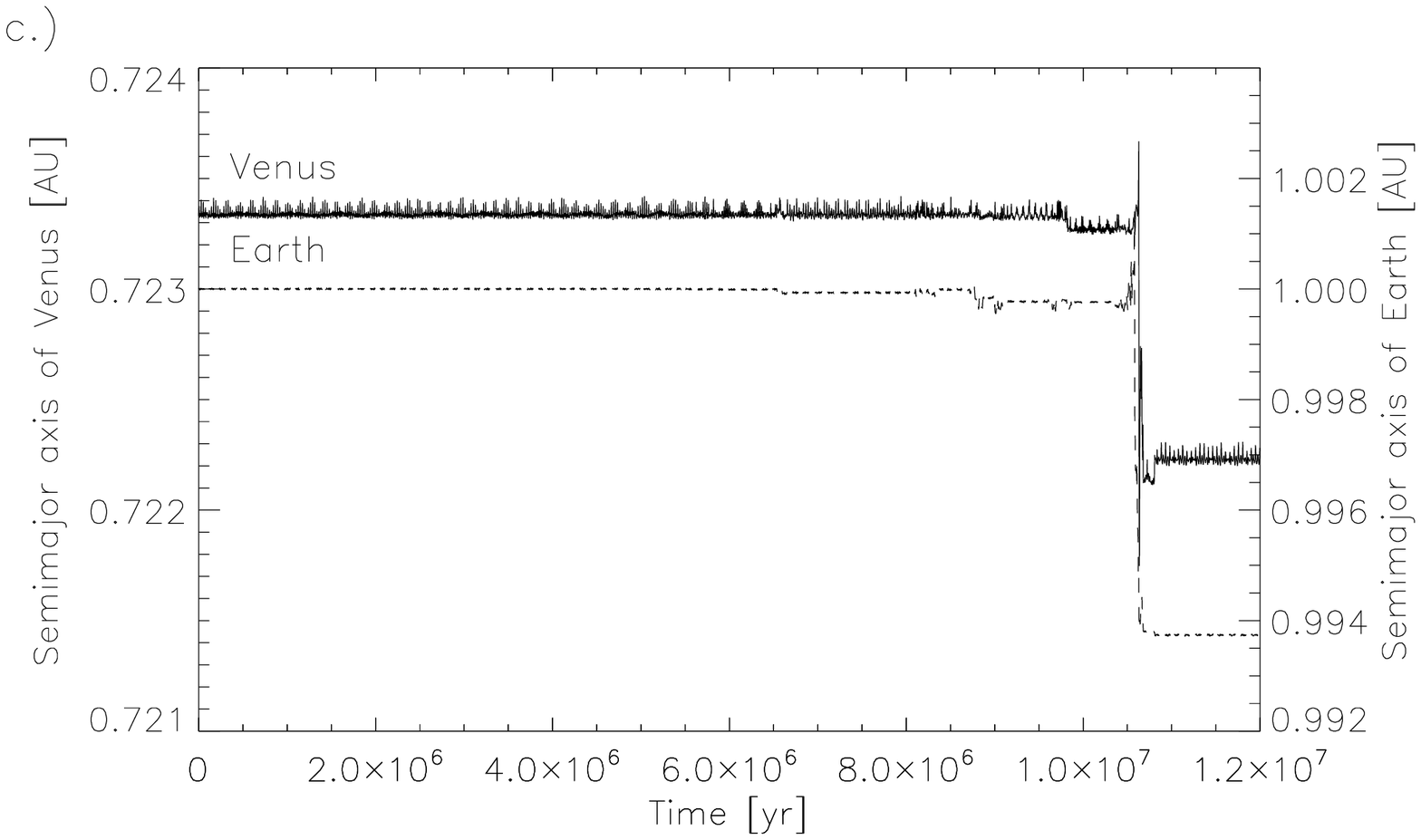}
\caption[]{a.) the inner planets' perihelia ($q$) and aphelia ($Q$) distances are shown as a function of 
time for the $\kappa_E=4.6$ simulation. b.) The eccentricity of Mars is depicted. Note that Mars' orbit is
excited to high eccentricity and it becomes Earth-crossing at $t \approx$ 10.5 Myr. c.) The semimajor
axes of the Earth and Venus are shown together.}
\label{fig6}
\end{figure}

Fig. \ref{fig6} shows the evolution of the perihelia ($q$) and aphelia ($Q$) distances of the inner 
planets for $\kappa_E = 4.6$, for which case we measured the shortest dynamical lifetime. A striking 
feature is that Mars' perihelion begins to decrease right at the start of the integration, as a consequence
of the steep increase of its eccentricity. At 250 000 yr Mars' $e$ reaches 0.16 then it decreases to 0.13,
around this value it oscillates for 1 Myrs. Conversely, $q$ and $Q$ of the Earth and Venus stay in a well 
defined zone. At $\sim$ 1.5 Myr, the perihelion of Mars begins to oscillate with quite a large amplitude. 
The center of this oscillation grows secularly and after 10.5 Myr Mars becomes an Earth-crosser, moreover 
it crosses the orbit of Venus too. As a consequence of the several close encounters with Earth; Mars escapes
from the system. This cascade mechanism is shown in Fig. \ref{fig6}c, where the semimajor axes of the 
Earth and Venus are plotted.

Because of the chaotic nature of the system, different dynamical lifetimes for the different $\kappa_E$ values 
were found. Out of the 21 integrations for 100 Myr, Mars escaped four times and the system decayed 
\footnote{By the decay of the system, we understand that the system has evolved into a different one, with a 
completely distinct configuration.}.
It must be stressed that the chosen length of the integration time is a problem, because it may be relatively 
short for such investigations. It is well-known that there exist orbits which are stable for a very long time
interval and then, all of a sudden they evolve into chaotic orbits. These 'sticky orbits' are common in nonlinear 
dynamical systems and therefore they exist also in planetary systems \citep{Dvorak1998,Jones2002}. This class of 
orbits are embedded in those region of the phase space, where regular and irregular orbits are close to each other.
We suspect that those systems which did not decay, may be 'sticky systems', and will be unstable at a much 
longer time.

According to the chaos theory the extent of this instability window is a function of the numerical resolution
(in our case the $\Delta \kappa_E$). Therefore it is not possible to find the exact size of this window, however 
with our resolution this instability window is in the interval 4.2 $< \kappa_E <$ 4.9

The surprising increase in the eccentricity of Mars may be a result of some secular resonances between 
the secular frequencies $g_j$ and $f_j$. Such phenomena were already found by Laskar in the Solar System, 
who reported that large and irregular variations can appear in the eccentricities and inclinations of 
the terrestrial planets, especially of Mercury and Mars on time scales of several Gyr \citep{Laskar96}.

The secular variations of the orbital elements of the planets are calculated by means of the Laplace-Lagrange 
theory. Using our {\sc MAPLE} application the $M_s^{(j)}$ and $L_s^{(j)}$ amplitudes, the $g_j$ and $f_j$ 
secular frequencies and $\beta_j$ and $\gamma_j$ were calculated in the $\kappa_E \in$ [4,6] interval 
on a very fine grid with a stepsize $\Delta\kappa_E =0.001$. In this region two frequencies, $f_2$ and 
$f_3$ have almost the same value (see Table \ref{tab:4.1}), while the other ones are well separated. 
In Table \ref{tab:4.1} the $f_5=0.0$ frequency is not included, and also the corresponding amplitude 
(2.8402 $\cdot 10 ^{-2}$) and the angular phase (106$^\circ$.17) were left out.

\begin{figure}
\centering
\includegraphics[width=1.0\linewidth,angle=0]{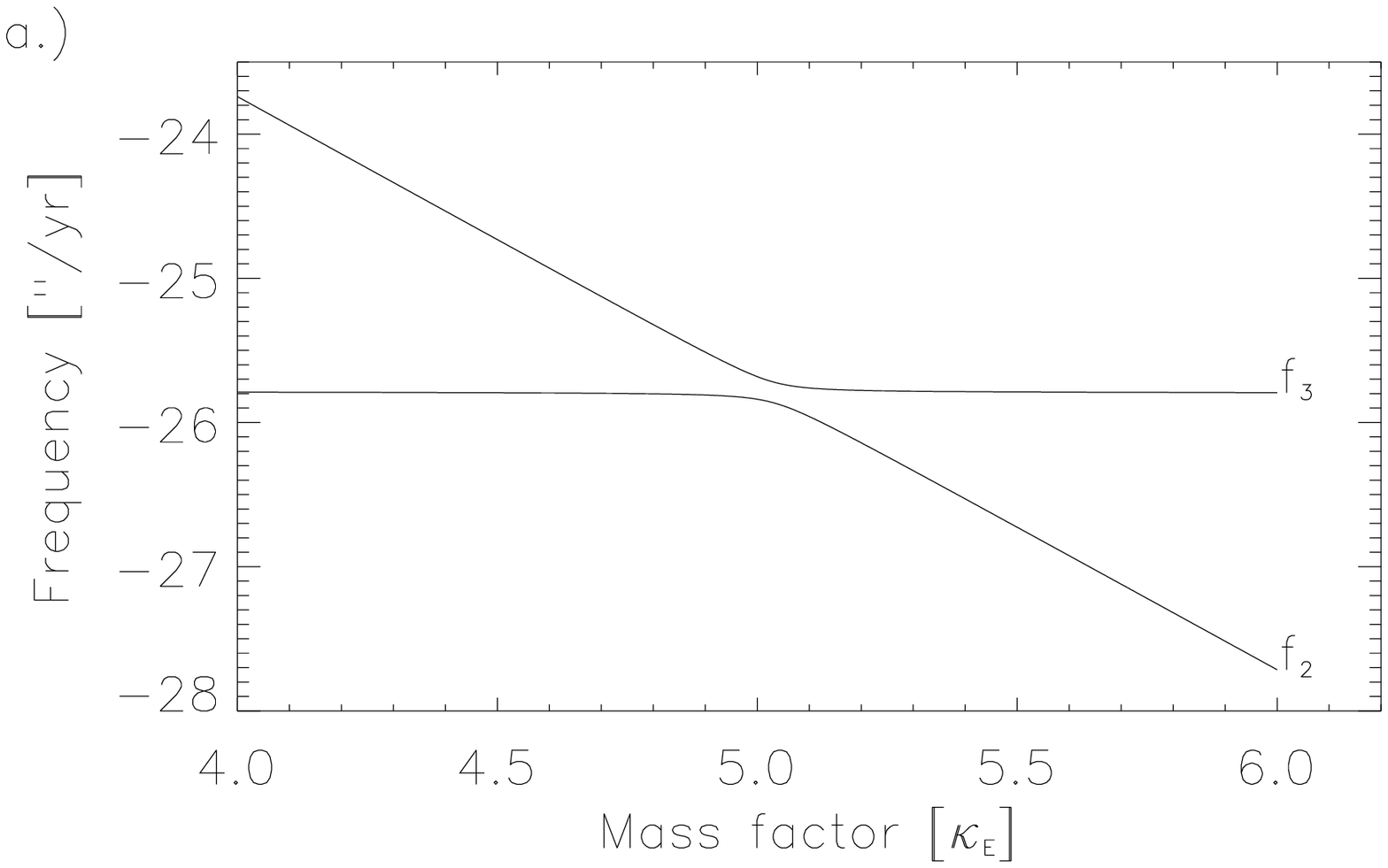}
\includegraphics[width=1.0\linewidth,angle=0]{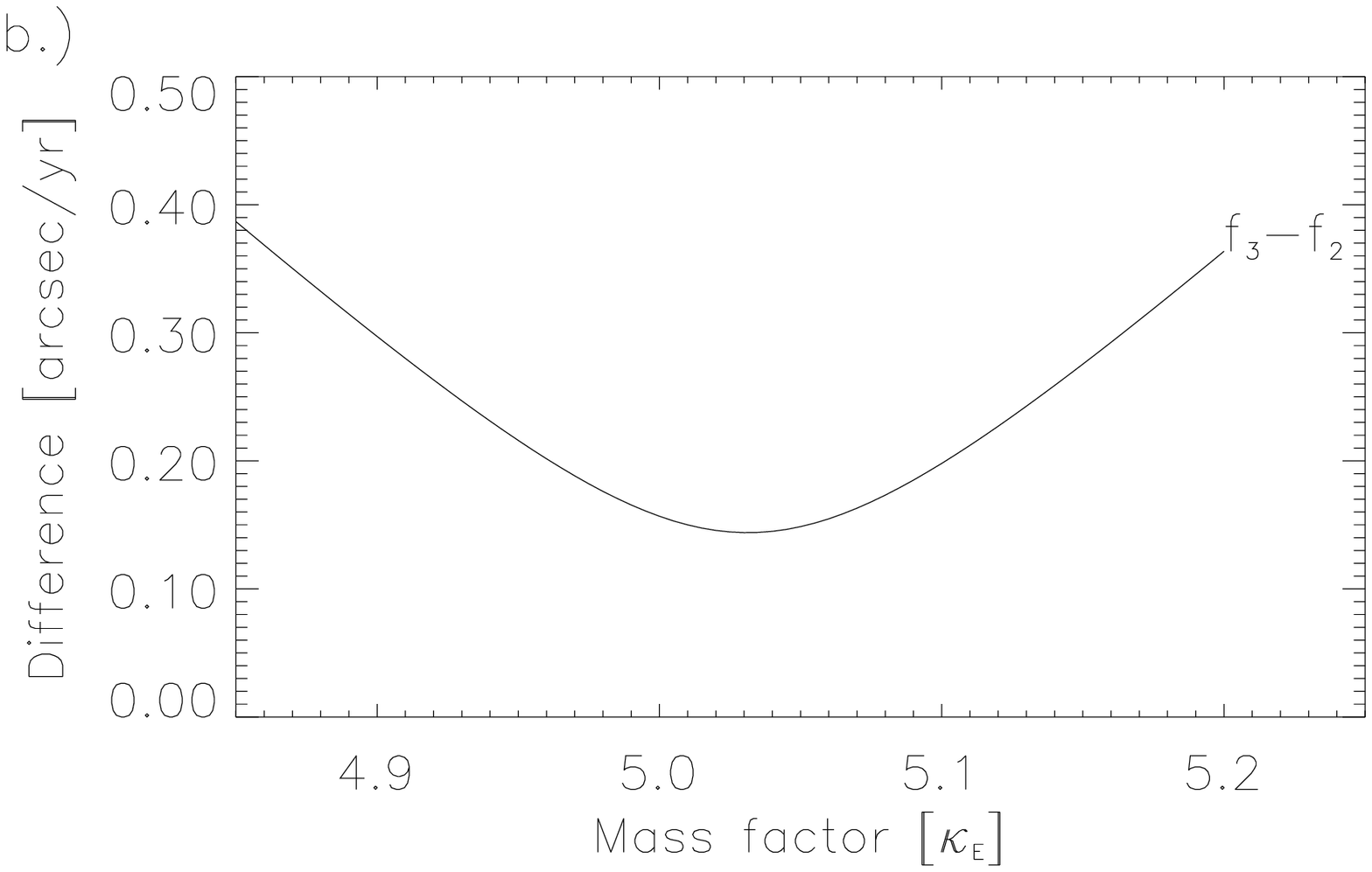}
\caption[]{On the a.) panel are the $f_2$ and $f_3$ secular frequencies, on the b.) panel their
difference around 5.0 are plotted versus the $\kappa_E$ mass factor.}
\label{fig7}
\end{figure}

To visualize the dependence of $f_2$ and $f_3$ on the mass factor, they are shown in Fig. \ref{fig7}a,
while in Fig. \ref{fig7}b their difference around $\kappa_E = 5$ is depicted. From Fig. \ref{fig7}b 
it is obvious, that the two curves do not intersect each other: the difference between the two frequencies 
are several orders of magnitudes greater than the accuracy of the computation, which was set to $10^{-10}$. 
The $\Delta f=f_2 - f_3$ difference reaches its smallest value for $\kappa_E=5.03$, $\Delta f=0.1439$ ''/yr, 
which is at the border of the aforementioned instability window.

A study of Table \ref{tab:4.1} shows that the largest amplitude are, in the solution for Mars, $L_3^{(2)}$ and $L_3^{(3)}$, 
for Jupiter $L_4^{(2)}$ and $L_4^{(3)}$ and for Saturn $L_5^{(2)}, L_5^{(3)}$. Accordingly 
the $f_2$ and $f_3$ frequencies, as was described in section 2.1, can be associated with Mars, Jupiter and Saturn.
The orbital plane of Mars therefore rotates together with those of Jupiter and Saturn, giving rise to chaotic 
behaviour. The equality of two apsidal or nodal rates is referred to in Solar System as a secular resonance.
In this case we have three secular resonances: $\dot \Omega_M \approx \dot\Omega_J$, $\dot\Omega_M \approx 
\dot\Omega_S$ and $\dot\Omega_J \approx \dot\Omega_S$. We suspect that these secular resonances are the
possible source of the observed chaos, and produce the instability window.

\begin{table*}
\centering
\begin{minipage}{140mm}
\caption{The $f_j$ secular frequencies, the $\gamma_j$ phases and the $L_s^{(j)}$ amplitudes of the model 
for $\kappa_E=5.03$ calculated using the Lagrange-Laplace theory. The $f_j$ are given in arcsec/yr and 
the $\gamma_j$s in degree.}
\centering
\begin{tabular}{|r| r r r r|}
\hline
$f_j$		&-47.381072 & \colorbox{sz}{-25.862186} & \colorbox{sz}{-25.718241} & -7.157383 \\
\hline
$\gamma_j$	& 76$^\circ$.92 & 306$^\circ$.01		& 128$^\circ$.89	& 297$^\circ$.61 \\
\hline
$L_s^{(j)}$ & & & & \\
Venus		& 5.2399 $\cdot 10 ^{-2}$ & 7.5778 $\cdot 10 ^{-3}$ & 1.0696 $\cdot 10 ^{-2}$ & 2.5795 $\cdot 10 ^{-2}$ \\
Earth		&-7.7184 $\cdot 10 ^{-3}$ & 3.9381 $\cdot 10 ^{-3}$ & 5.9598 $\cdot 10 ^{-3}$ & 2.3939 $\cdot 10 ^{-2}$ \\
Mars		& 2.4032 $\cdot 10 ^{-3}$ &-4.2737 $\cdot 10 ^{-1}$ &-4.3543 $\cdot 10 ^{-1}$ & 1.3348 $\cdot 10 ^{-2}$ \\
Jupiter		& 4.5852 $\cdot 10 ^{-6}$ & 3.1885 $\cdot 10 ^{-3}$ &-3.1088 $\cdot 10 ^{-3}$ &-1.2056 $\cdot 10 ^{-4}$ \\
Saturn		&-2.2595 $\cdot 10 ^{-6}$ &-7.7642 $\cdot 10 ^{-3}$ & 7.7424 $\cdot 10 ^{-3}$ &-1.8405 $\cdot 10 ^{-4}$ \\
\hline
\end{tabular}
\label{tab:4.1}
\end{minipage}
\end{table*}

\subsection{The $30 \leq \kappa_E \leq 600$ region}

\begin{figure}
\centerline{
\includegraphics[width=1.0\linewidth,angle=0]{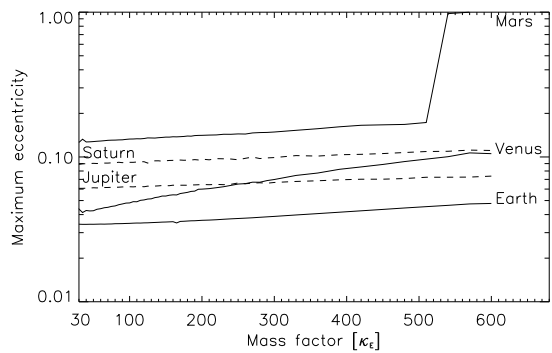}}
\caption{Results of the MEM in the $30 \leq \kappa_E \leq 600$ region. Note, that
the $y$ axis is logarithmic.}
\label{fig8}
\end{figure}

Fig. \ref{fig8} summarizes the results of the MEM in the $30 \leq \kappa_E \leq 600$ region. As one can 
clearly see from Fig. \ref{fig8} the MEs of the planets gradually increase. None of the curves show any 
peaks, consequently in this region no instability window was found. These numerical results are in line
with the results of the linear theory. In this region each of the secular frequencies have quite different 
values. Beyond $\kappa_E \ge 540$ Mars escaped in all simulations. The MEs of the Earth, Jupiter and Saturn 
behave very similarly and we may say that from $\kappa_E = 100$ on the system is dominated by these planets 
and Venus and Mars may be considered as quasiasteroids.

Generally the semimajor axes of the inner planets are confined in a narrow stripe in the order of 
$10^{-4}$--$10^{-3}$ AU. These stripes are defined by a very short quasiperiodic oscillation around
the mean values of the planets' semimajor axes. As the mass factor increases the amplitudes of the
oscillations grow, the width of these stripes slightly broadens.

\begin{figure}
\centering
\includegraphics[width=1.0\linewidth,angle=0]{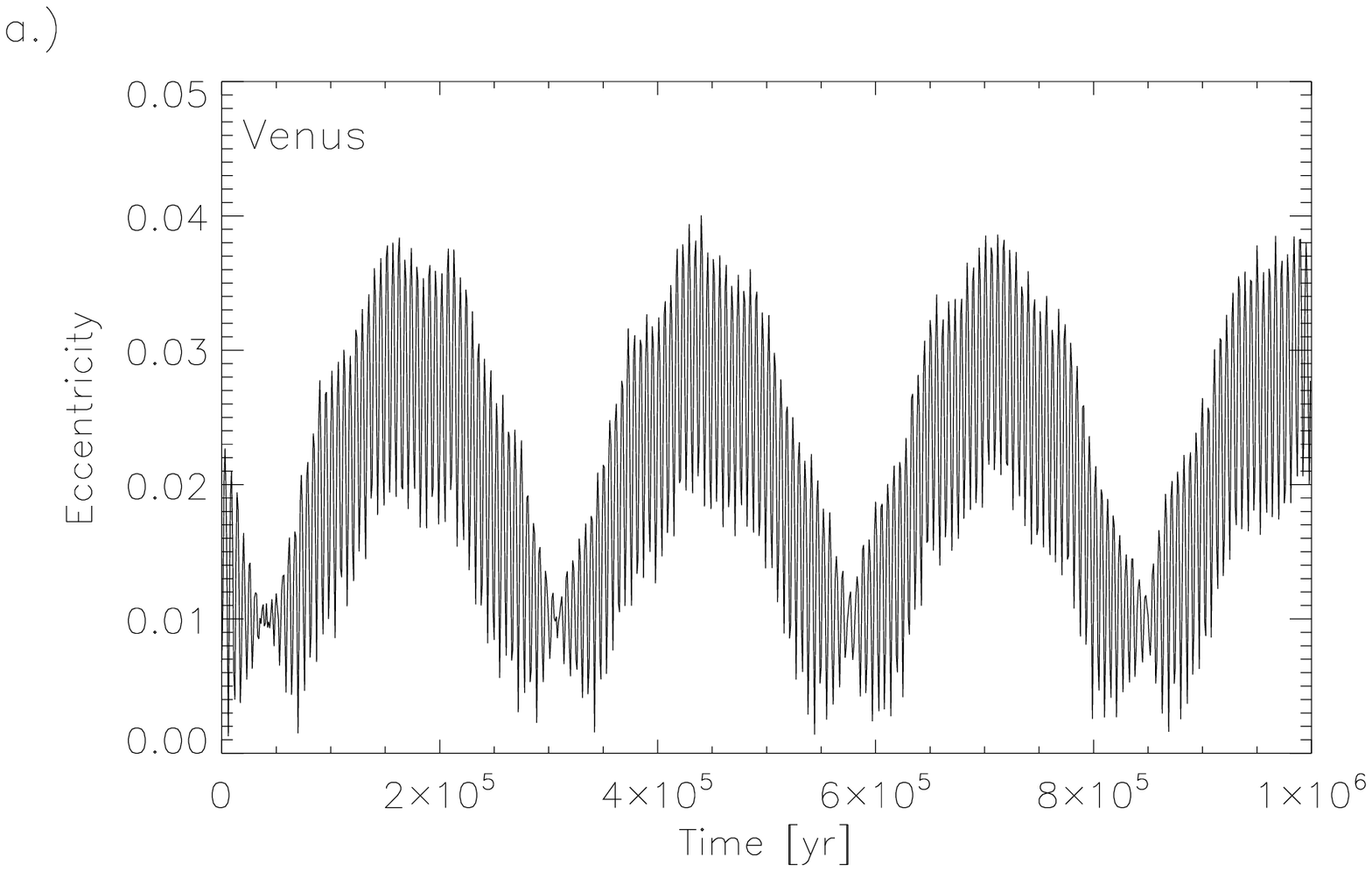}
\includegraphics[width=1.0\linewidth,angle=0]{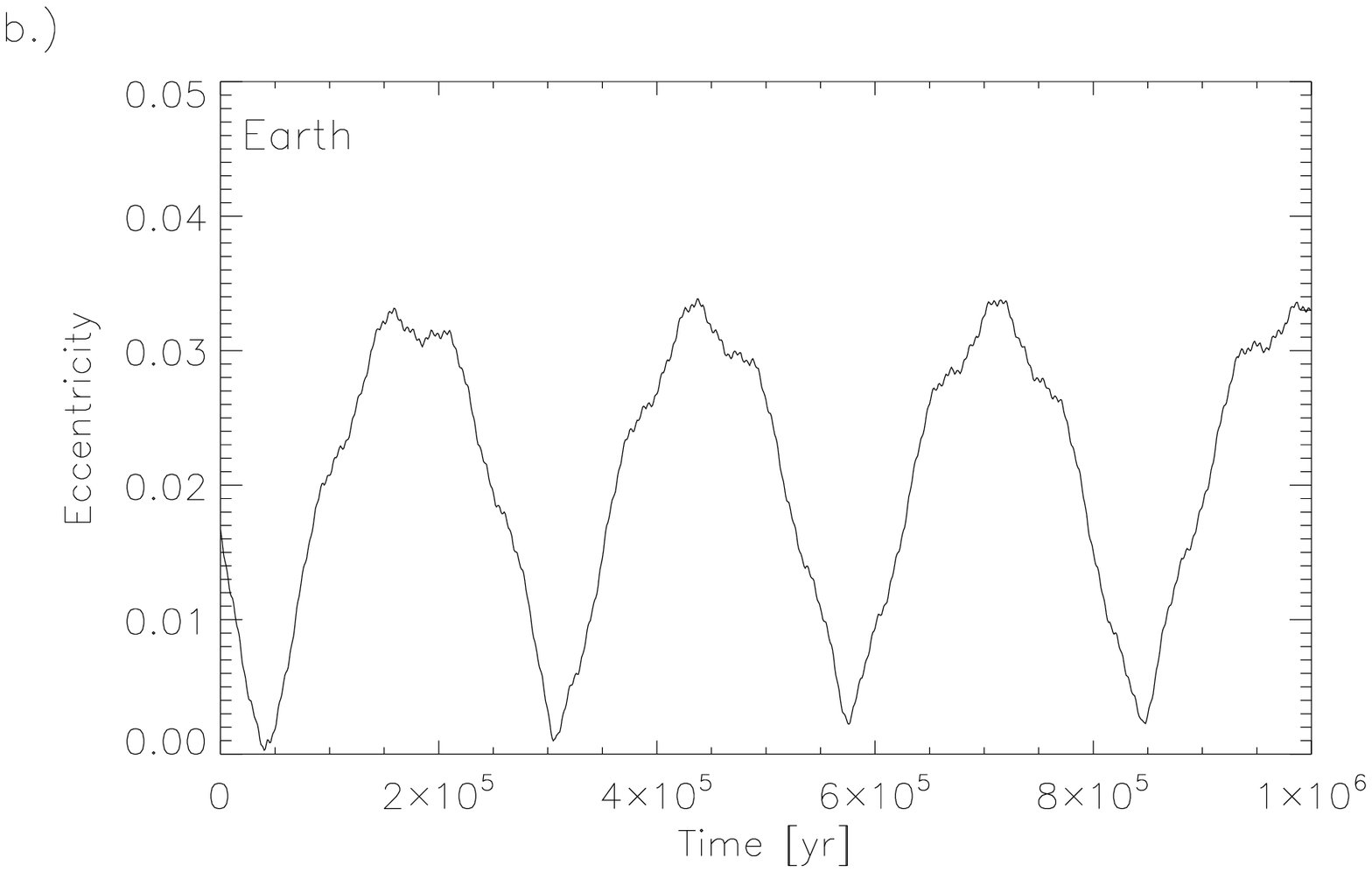}
\includegraphics[width=1.0\linewidth,angle=0]{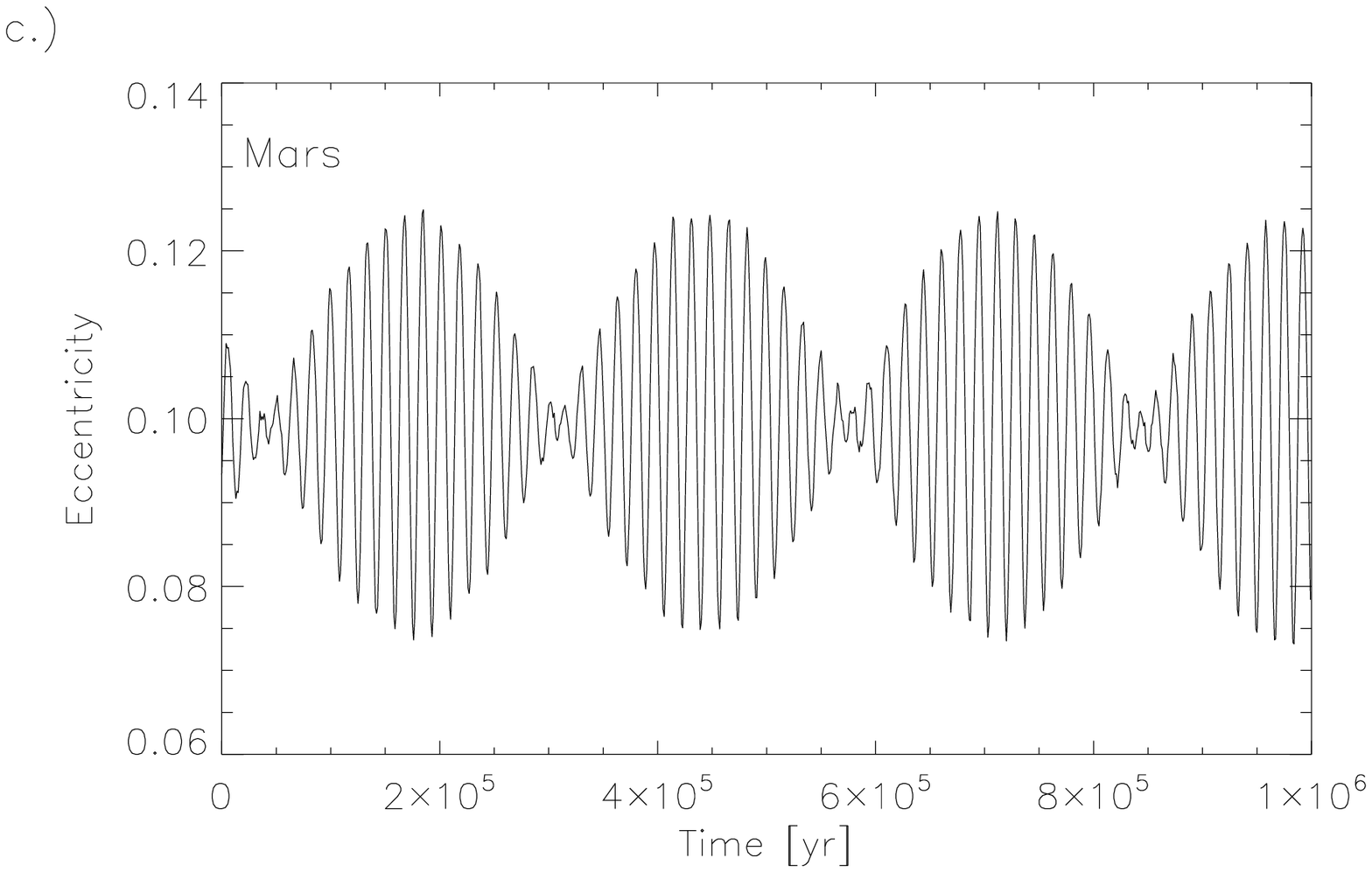}
\caption{The eccentricities of the Venus, Earth and Mars (from top to bottom) for $\kappa_E = 30$
over a 1 million year time-span.}
\label{fig9}
\end{figure}

The typical evolution of the eccentricity of the inner planets are shown in Fig. \ref{fig9}. The 
character of the curves are similar to Fig. \ref{fig4}c, only the periods are shorter.

Since Venus, Earth and Mars have the same period in their eccentricities -- just like the Jupiter 
and Saturn pair does -- the system is separated into two subsystems: one consisting of Venus, Earth 
and Mars, and the other Jupiter and Saturn. This behaviour was already observed for smaller 
mass factor values. The strong coupling of the eccentricities of the two giant planets remains unbroken 
for all $\kappa_E$. The eccentricities of Venus and Mars are mainly determined by that of the Earth,
which is the superposition of a long period ($T_l \sim 2.5\cdot10^{5}$ yr) variation with amplitude 0.035
and several shorter period ($\sim 10^{4}$ yr) and smaller amplitude variations (see Fig. \ref{fig9}b). 
In the case of Venus, upon this long period variation several very short periods are superimposed (see 
Fig. \ref{fig9}a). On the contrary the eccentricity of Mars flickers with a period $T_l$ around 
0.1 (see Fig. \ref{fig9}c).

\begin{figure}
\centering
\includegraphics[width=1.0\linewidth,angle=0]{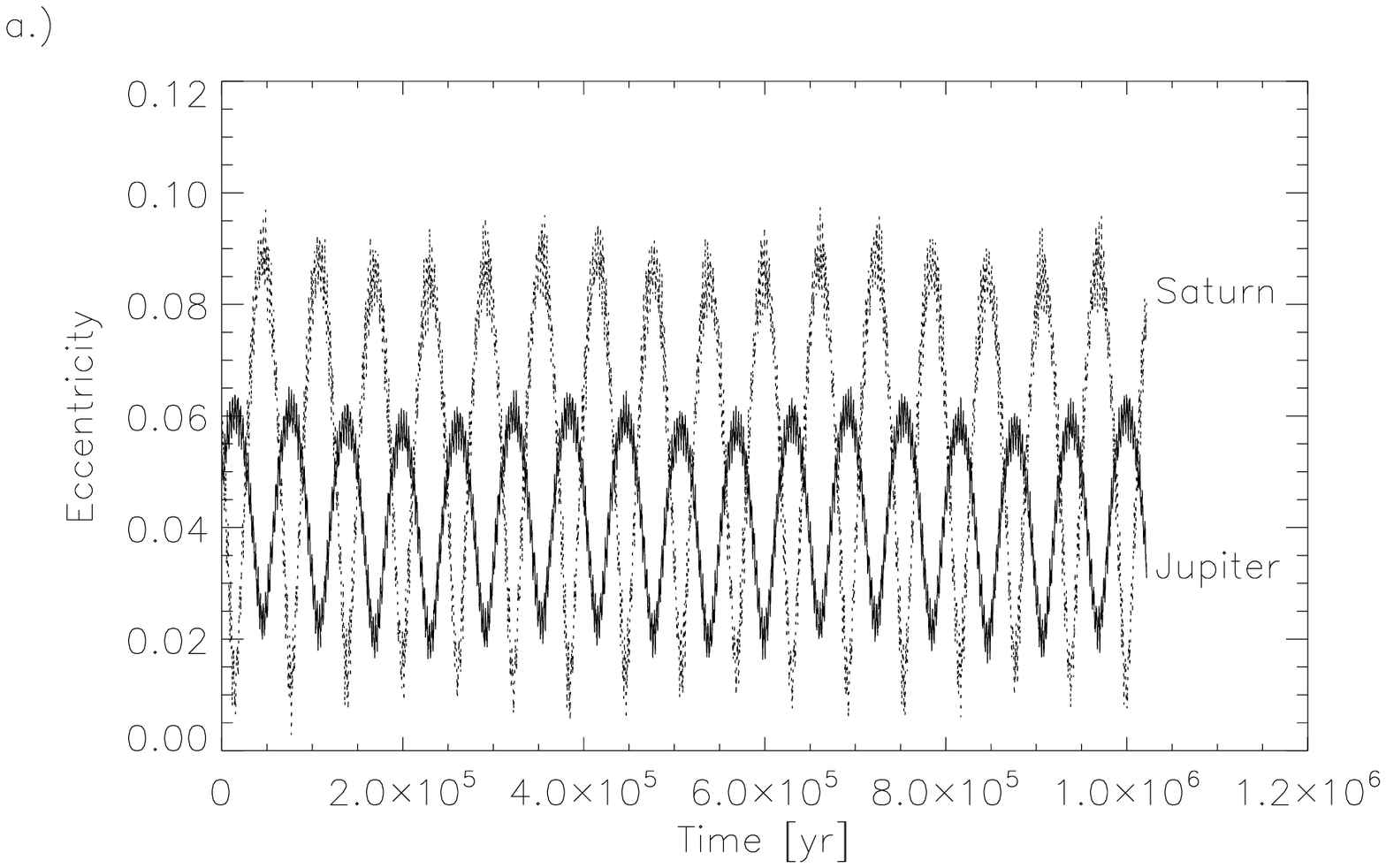}
\includegraphics[width=1.0\linewidth,angle=0]{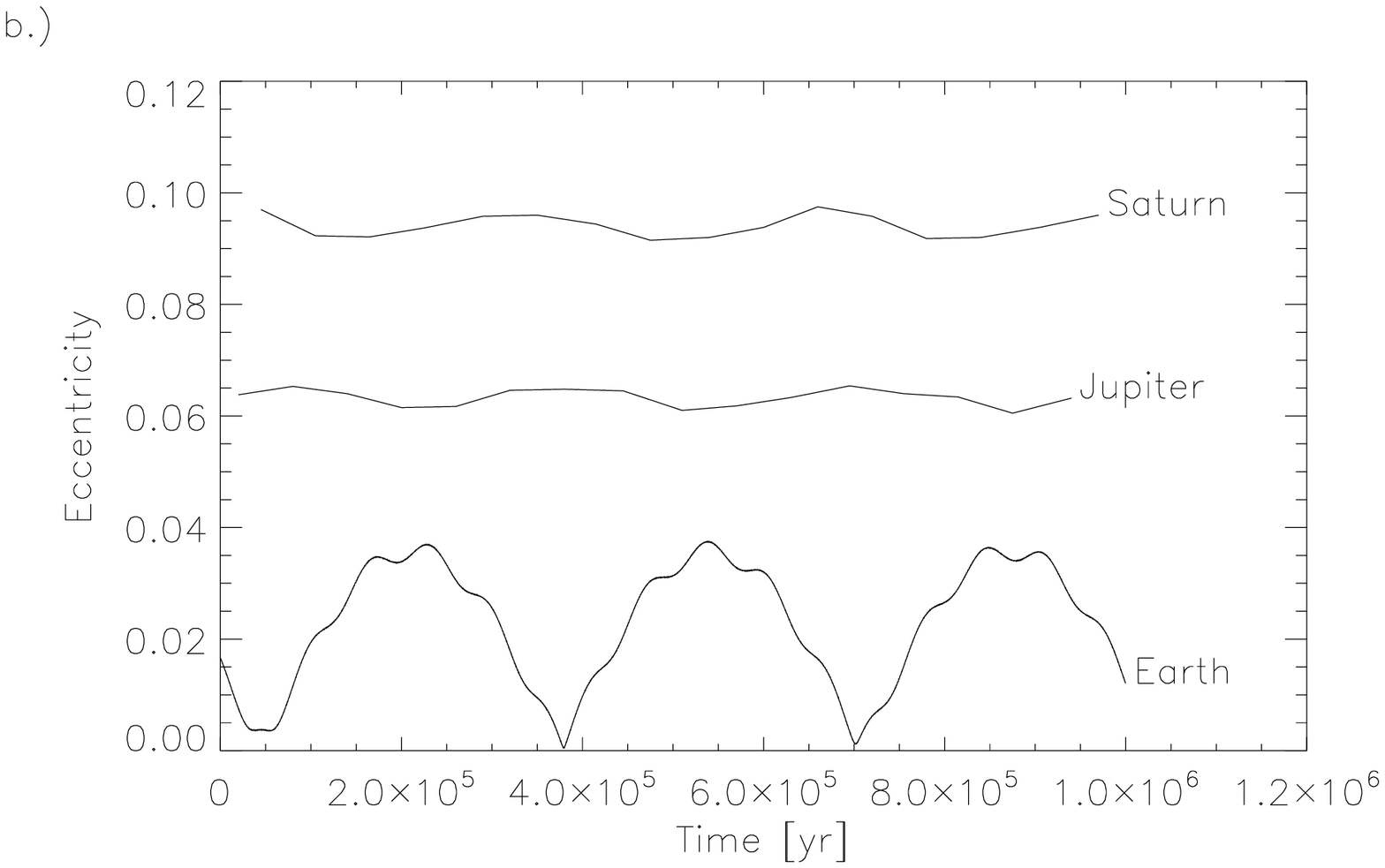}
\caption{a.) The time development of the eccentricities of Jupiter and Saturn. b.) The eccentricity of the Earth 
and the local maximums of Jupiter's and Saturn's eccentricities are shown for $\kappa_E = 250$. The coupling 
between the eccentricities can be well observed.}
\label{fig10}
\end{figure}

As the mass factor increases the long period variations in the eccentricities of Jupiter and Saturn are more and 
more influenced by the Earth. In order to conserve the total angular momentum of the system the centers 
of oscillations of Jupiter's and Saturn's eccentricities are anticorrelated with the eccentricity of the 
Earth (see Fig. \ref{fig10}b). In spite of the large perturbations from the Earth, the coupling between 
the motion of Jupiter and Saturn are still very well determined (see Fig. \ref{fig10}a). This result is 
also supported by our previous study \cite{Dvorak2002}, and therefore we may say that the dynamical coupling 
of Jupiter and Saturn essentially determines the dynamics of the Solar System's bodies.

The time developments of the inclinations are very similar to those of the eccentricities:
\begin{itemize}
\item the variations of Jupiter's and Saturn's inclinations are coupled,
\item the inclinations of Venus and Mars are primarily determined by that of the Earth,
\item with growing $\kappa_E$ Jupiter's and Saturn's inclinations are increasingly influenced by the Earth.
\end{itemize}

\begin{figure}
\centering
\includegraphics[width=1.0\linewidth,angle=0]{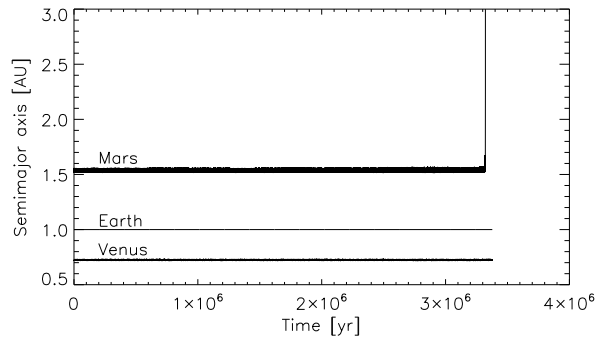}
\caption{This plot shows the escape of Mars for $\kappa_E$ = 540.}
\label{fig11}
\end{figure}

When $\kappa_E = 540$ Mars' eccentricity had grown secularly and at $T_e \approx$ 3.4 Myr, Mars
passed near by Earth, which ejected the planet from the system (see Fig. \ref{fig11}). In this case, 
we did not observe orbit-crossing. 
However the minimum distance between Earth and Mars was 0.164 AU, which is about twice the Hill radius of 
the massive "Earth" $R_H(\kappa_E = 540) = 0.081$ AU. A cataclysmic outcome is therefore to be expected 
\citep{Jones2005}. In the systems with $\kappa_E$ = 570 and 600 Mars was ejected after a sequence of 
orbit-crossing with Earth at $T_e \approx 4.35 \times 10^6$ and at $T_e \approx 5.51 \times 10^6$ years,
respectively.

\begin{figure}
\centering
\includegraphics[width=1.0\linewidth,angle=0]{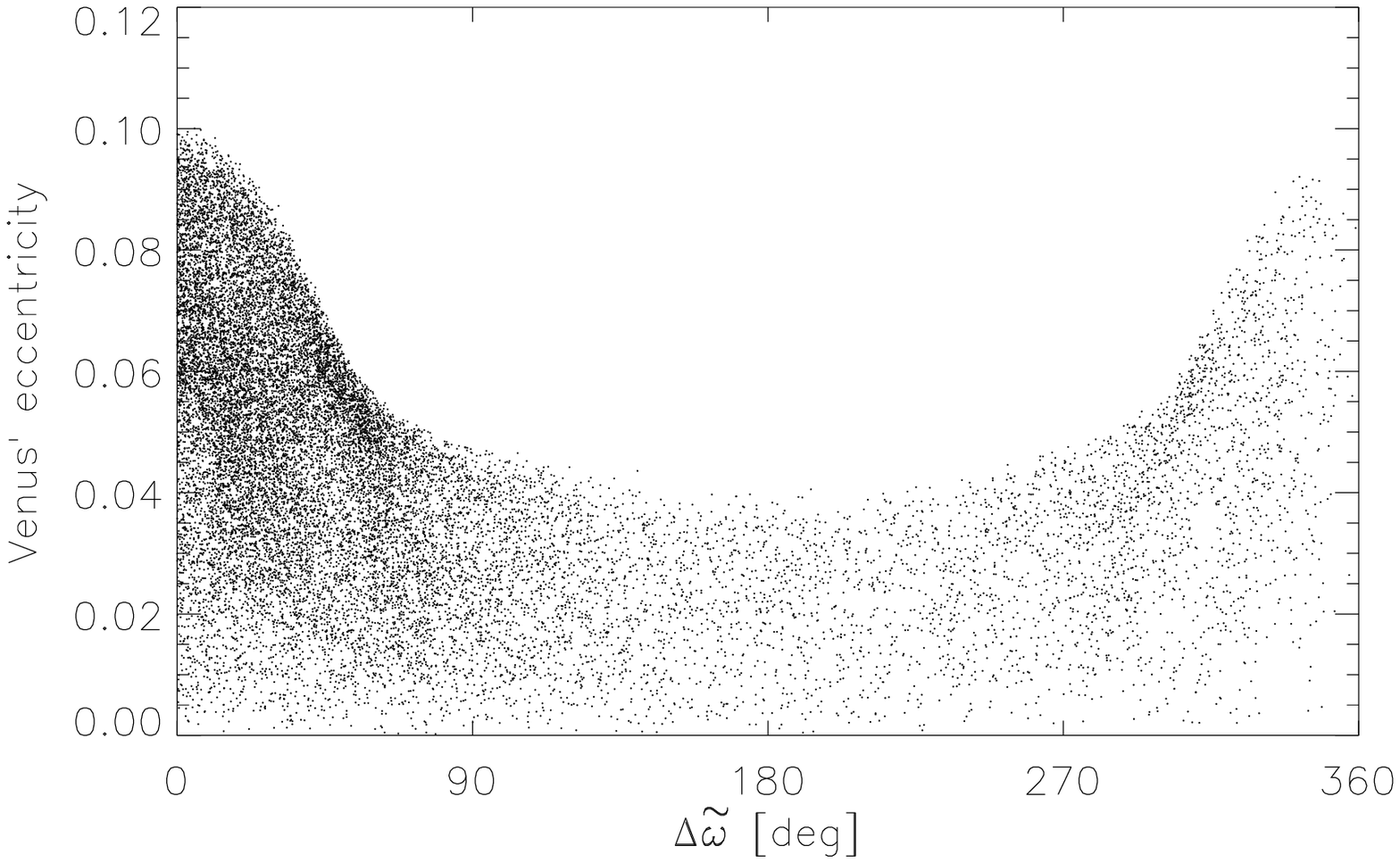}
\caption{This plot shows the relationship between Venus' eccentricity and $\Delta \varpi$ during
20 Myrs for $\kappa_E$ = 540.}
\label{fig12}
\end{figure}

The stability of Venus is an intriguing property of the systems. In each of our simulations the orbital
elements of Venus did not show any sign of chaos. The Earth and Venus are prevented from close encounters 
by a coupling of Venus' eccentricity and $\Delta \varpi$; $\Delta \varpi$ is the difference between the
longitude of perihelion of the Earth and Venus. Whenever the perihelion of Earth conjunctions with the
aphelion of Venus ($\Delta \varpi = 180^\circ$), the eccentricity of Venus is around its minimum (see 
Fig. \ref{fig12}), maximizing the distance between the two orbits. A similar protection mechanism also 
occurs with real Mars and the asteroid Pallas. This kind of coupling was also observed by \cite{Jones2002}
for an 'Earth' in the 47 UMa system.

\section{Discussion}

We studied the dynamics of a simplified dynamical model of the Solar System where we included the 
three terrestrial planets Venus, Earth and Mars and the gas giants Jupiter and Saturn. This model 
was already studied in detail \citep{Dvorak2002} when the masses of all terrestrial planets were 
uniformly increased. It turned out that the different systems remained stable for 10 Myrs
up to large mass factor values. As a continuation of this work the same model with a more massive Earth
was studied and signs of chaotic behaviour were reported in \cite{Dvorak2005}. In this article a
detailed exploration of the different $S^3$ setups is presented, and the existence of an 
instability window is revealed. As an important byproduct the plethora of the integrations can
serve as a general model of exoplanetary systems with two massive planets close to the 5:2 mean 
motion resonance on low eccentric orbits for comparable mass ratios of the giant planets.

We investigated over 100 $S^3$ models where we increased the mass of the Earth by a mass factor $\kappa_E$
between $1 \le \kappa_E \le 600$. These new systems have been integrated using extensive numerical integrations
for 20 million years (for selected systems up to 100 million years) to find out the effect of a very massive
Earth on the stability of the whole system on one hand; on the other hand the model now can serve as example 
of exoplanetary systems of two or three massive planets.

It turned out, that even when the Earth had Jupiter's mass ($\kappa_E \approx 300$) and beyond, the system 
was still stable, but when $\kappa_E \in (4.2\,,4.9)$, the motion of Mars become chaotic. In this instability
window Mars' orbital 
eccentricity finally reached values which led to close encounters of Mars with the Earth, and even with 
Venus. After a sequence of close encounters Mars escaped within some millions of years. Using the results of the
Laplace-Lagrange secular theory we found secular resonances acting between the motions of the nodes of Mars,
Jupiter and Saturn. These secular resonances give rise to strong chaos, which is the cause of the appearance
of the instability window, and eventually the escape of Mars. 

We also found an interesting coupling of Venus' $e$ and $\Delta \varpi$, which protect Venus
form a close encounter with Earth. This mechanism was observed in the Solar System and in 47 UMa
with a hypothetical Earth \citep{Jones2005,Jones2002}.

According to these results, the stability of the Solar System depends on the masses of the planets, 
and small changes in these parameters may result in a different dynamical evolution of the planetary system. 
No other instability window in $\kappa_E$ were found; first results of additional computations where we increased 
the masses of Venus and separately also of Mars, showed signs of chaotic motions for some windows in $\kappa$ too, 
but a detailed study of these two other cases of a modified $S^3$ is in preparation. There we also intend to give 
a detailed comparison of the three systems, namely with a massive Venus, a massive Earth and a massive Mars.

Finally we note that all these models may be used as dynamical reference models for a better understanding 
of the stability of orbits in extrasolar planetary systems.

\section*{Acknowledgments}
We would like to thank Professor B. Érdi for a critical reading of the original version of the paper 
and also the anonymous referee for several comments which greatly improved the clarity of the paper.
We thank the Wissenschaftlich-Technisches Abkommen \"Osterreich-Ungarn Projekt A12-2004, and
the Hungarian Scientific Research Fund T043739. All numerical integrations were accomplished 
on the NIIDP (National Information Infrastructure Development Program) supercomputer in Hungary.

\label{lastpage}
\end{document}